\documentclass[fleqn,10pt]{wlscirep}

\usepackage{graphicx}

\usepackage[utf8]{inputenc}
\usepackage[T1]{fontenc}

\title{COVID-19 denialism in Brazil: a multifactor study}

\author[1,*]{Tarcísio M. Rocha Filho}
\author[2]{Magda L. Lucio}
\author[3]{Fulvio A. Scorza}
\author[4]{Marcelo A. Moret}

\affil[1]{ICP-IF, Universidade de Brasília, Campus Universitário Darcy Ribeiro, Asa Norte, Brasília, DF, Brazil}
\affil[2]{FACE, Universidade de Brasília, Campus Universitário Darcy Ribeiro, Asa Norte,
          Brasília, DF, Brazil}
\affil[3]{Escola Paulista de Medicina, Universidade Federal de São Paulo, São Paulo – Brazil}
\affil[4]{SENAI-CIMATEC, Salvador, BA, Brazil and UNEB, Salvador, BA, Brazil}

\affil[*]{marciano@fis.unb.br}

\begin{abstract}
We discuss the relationships between the outcome of the COVID-19 pandemic in Brazil at the municipal level
and different health, social, demographic, and economic indices. We obtain significant correlations
between the data gathered for each municipalitiy
and the proportion of cases and deaths by COVID-19 and the results by municipality of the 2018 Brazilian
presidential election. We obtain different estimates for the number of deaths
caused by central government denialism of scientific facts and measures for mitigation of the pandemic and
its the historical, economic, and social roots.
\end{abstract}


\begin{document}

\flushbottom
\maketitle

\section{Introduction}
\label{secint}

Amidst the different issues resulting from the policies implemented by different governments
the capacity to deal with crisis is certainly the one with the most overreaching consequences.
This was staunchly demonstrated with the advent of the COVID-19 pandemic, caused by the SARS-CoV-2 virus.
According to official reports~\cite{johnhopkins}, since its first detection in Wuhan (China) in December
2019 to the present date, the virus has been responsible for six and a half million deaths and 615 million cases.
Although these figures are sadly impressive in and of themselves, the real numbers are much worse,
with cases significantly under-reported~\cite{jrah} and under-reported deaths proportionately smaller
but still significant~\cite{Noorden2022}. The estimated real number of deaths in the world is approximately
18 million~\cite{IHM2022}, with under-reporting varying significantly among countries and regions.
This only shows that the burden of COVID-19 is by far the worst planetary health crisis in more than a century.

A recent report by the Lancet Commission on lessons for the future from the
COVID-19 pandemic~\cite{Commission2022} addresses both the correct measures
and failures in the mitigation of the COVID-19 pandemic and its social and economic consequences.
Among different analyses, the report points out that most governments and the World Health Organization
were too slow to react to the new pandemic; that many mitigation measures were hindered
by wide sectors of the population; and that trust in communities acted as an important fighting
tool against the pandemic.
Some authors have addressed the consequences of pandemic mitigation policies according to government choices
and public discourse.
Based on surveys from the initial stages of the pandemic, Pickup and collaborators
discussed the effects of political partisanship on attitudes and perceptions of government policies
in mitigating the pandemic in the US and Canadai~\cite{Pickup2020},
with evidence that partisanship guided the assessment of central government policies
against COVID-19 in both countries. Bennouna et al.\ studied the great variations of COVID-19 mitigation policies
in Brazil, Mexico, and the US and the importance of central (presidential) and state (governors) coordination,
or its absence, in the adoption of public policies~\cite{Bennouna2021}. The effect of political partisanship on
adherence to social distancing in each US state was studied using mobility statistics and found to be closely related
to the respective governor's political party, with recommendations being more effective in Democratic than in Republican
counties~\cite{Grossman2020}. A comparison of public health emergency measures and social policiesD in
demonstrated that their simultaneous implementation
succeeded in confronting the pandemic, as was the case in Germany; that social policies without
associated health interventions failed to result in an effective mitigation, as occurred in Brazil and the US; while
in India, public health policies simply failed due to the absence of social interventions~\cite{Greer2021}.
Today, among different public policies, the most effective against COVID-19 is mass vaccination, which, if properly
planned, can significantly reduce the burden of lost lives~\cite{Optim2021}.
Among the many difficulties faced in fighting the pandemic, denial of disease severity
and of scientific knowledge have a significant impact on the burden of health services and on the outcomes of the pandemic~\cite{Suarez2022}

Denialism in Brazil has been a major issue in mitigating the pandemic, with frequent questioning of mask wearing and
social distancing efficacy, the proposal of early treatment without any scientific evidence to corroborate its use~\cite{Malta2021}
and even backed by a number of physicians~\cite{Ferigato2020}, and denial of vaccine efficacy and safety~\cite{Hallal2021}.
Its impacts in numbers are difficult to estimate, but its effects are quite visible. For example, the number of COVID-19
deaths in the Brazilian public health system from January 1, 2021, to March 23, 2022, was significantly higher in non-fully
vaccinated individuals than in fully vaccinated individuals~\cite{Murari2022}, a consequence of ideological components
of COVID-19 denial resulting from erratic and often misleading speeches from the far-right Brazilian President and
central government authorities~\cite{Neves2022,Kibuuka2020}.
Indeed, the central government never issued mask mandates or any form of lockdown
in order to prevent the virus' spread, and there was no systematic testing policy.
Instead, state and municipal authorities implemented these measures, with varying degrees of success~\cite{Jorge2021,Rocha2021}.
These studies put forward the need for strong coordinated actions between different levels of government for
the simultaneous adoption of social and health policies for an efficient mitigation of the current
and future pandemics~\cite{Bergeijk2022}.

Here we investigate the relations between series of social, economic, and demographic indicators with
the electoral results for the 2018 second round of the Brazilian presidential election,
and how they determined the outcomes of the COVID-19 pandemic in Brazil, and
obtain estimates for the number of deaths related to COVID-19 denialism.
The current study relies on a large and representative set of data,
significantly expands on previous studies focusing on the COVID-19 and election relationships,
and discusses the picture that emerges from such analysis as well as its social and historical roots.

\section{Data}

The Federate Republic of Brazil is divided into 26 states plus the Federal District, and each state
is divided into 5570 municipalities, with a population ranging from $10\,004$ to $12\,228\,009$
(2022 estimates) and an area from $3,656,\rm km^2$ for Santa Cruz de Minas in the state of Minas Gerais
up to $159\,533 \rm km^2$ for Altamira in the state of Pará (northern region).
A municipality usually, but not always, corresponds to a city or a small conglomerate of cities.
The following data and respective sources were used in our analysis:
\begin{itemize}
\item Total number of COVID-19 cases and deaths by Brazilian municipality~\cite{minsaude}.
\item Number of COVID-19 vaccine by dose and type as a function of time for each municipality~\cite{openDataSUS}.
\item Number of votes received by each candidate at the second round of the 2018 Brazilian election~\cite{tse}.
\item Budget from the federal government for the public health structure received during the year of 2021~\cite{openDataSUS}.
\end{itemize}
The following social, economic and demographic data for the year of 2010
from the last official completed Census in Brazil:
\begin{itemize}
\item Illiteracy rate by municipality, for individual with 18 years of age or more~\cite{ipea}.
\item Proportion of individuals in each municipality with 25 years or more having completed their primary education~\cite{ipea}.
\item GINI index by municipality~\cite{ipea}.
\item Percentage of extremely poor individuals~\cite{ipea}.
\item Average income in each municipality~\cite{ipea}.
\item Human Development Index in each municipality~\cite{ipea}.
\item Population by self-declared race (skin color) in each municipality~\cite{dadosraca}.
\end{itemize}
Although the census data is a decade old as the 2020 census was postponed due to the pandemic, 
we do not expect a significant change in the rates and proportions used here, but we recognize it as
a source of error in the analysis to be presented below.
The recent estimates for the population in each municipality:
\begin{itemize}
\item Estimated population in 2022 in each municipality~\cite{populacao}.
\end{itemize}
The list of all variable considered in the correlation analysis is summarized in Table~\ref{tab1}.

\begin{table}[ht]
        \begin{center}
\begin{tabular}{cl}
        \hline\hline
        Data & Variable\\
        \hline
        Election result & $E_r$\\
        COVID-19 attack rate & $A_r$\\
        COVID-19 mortality rate & $M_r$\\
        Vaccine coverage (2 doses) in the population & $V_c$\\
        Adult illiteracy rate & $I_r$\\
        GINI coefficient & $G_i$\\
        Percentage of population in extreme poverty & $P_p$\\
        Average income & $A_I$\\
	Human development index & {\rm\it HDI}\\
        Per capita public health budget & $H_b$\\
        Percentage of black and indigenous population & $BI$ \\
        \hline
\end{tabular}
        \end{center}
        \caption{Variables considered in the analysis.\label{tab1}}
\end{table}

\section{Methods}
\label{secmeth}

In order to determine the intrinsic relations between each pair of variables listed in Table~\ref{tab1},
we use the Spearman rank-order correlation $r_s(A,B)$ between two ordered series
$A=(A_1,$ \ldots $,A_{N_{\rm data}})$ and $B=(B_1,\ldots,B_{N_{\rm data}})$ each composed by $N_{\rm data}$ values.
It is given by the Pearson correlation between their rank values,
and for the special case that all ranks are distinct by~\cite{myers}:
\begin{equation}
        r_s[A,B]=1-6\frac{\sum_{i=1}^{N_{\rm data}}d_i^2}{N_{\rm data}(N_d^2-1)},
        \label{spearmandef}
\end{equation}
with $d_i$ the difference in paired ranks of the two data sets $A$ and $B$ given by the difference in position of the $i$-th
data point in the two data sets in ascending order. The Spearman correlation satisfies $-1\geq r_s\geq1$ and
is a measure for how two variables are monotonically related, by an increasing or decreasing relation if $r_s>0$ or $r_s<0$,
respectively.

The results for the 2018 presidential election in Brazil (second round) are represented in each municipality by an index
defined as
\begin{equation}
        E_r=\frac{{\cal V}_1-{\cal V}_2}{P_m},
        \label{defEr}
\end{equation}
where ${\cal V}_1$ and ${\cal V}_2$ are the votes received by the left-wing and
the far-right candidates, Fernando Haddad and  Jair Bolsonaro, respectively,
and $P_m$ is the population in the given municipality.

\section{Results}

\subsection{COVID-19 outcomes and social and economic indices}

In order to have statistical significance for the COVID-19 data
we considered in the present analysis the $3079$ municipalities with an
estimated population of at least $10\,000$.

\begin{figure}[ht]
\begin{center}
        \scalebox{0.45}{\includegraphics{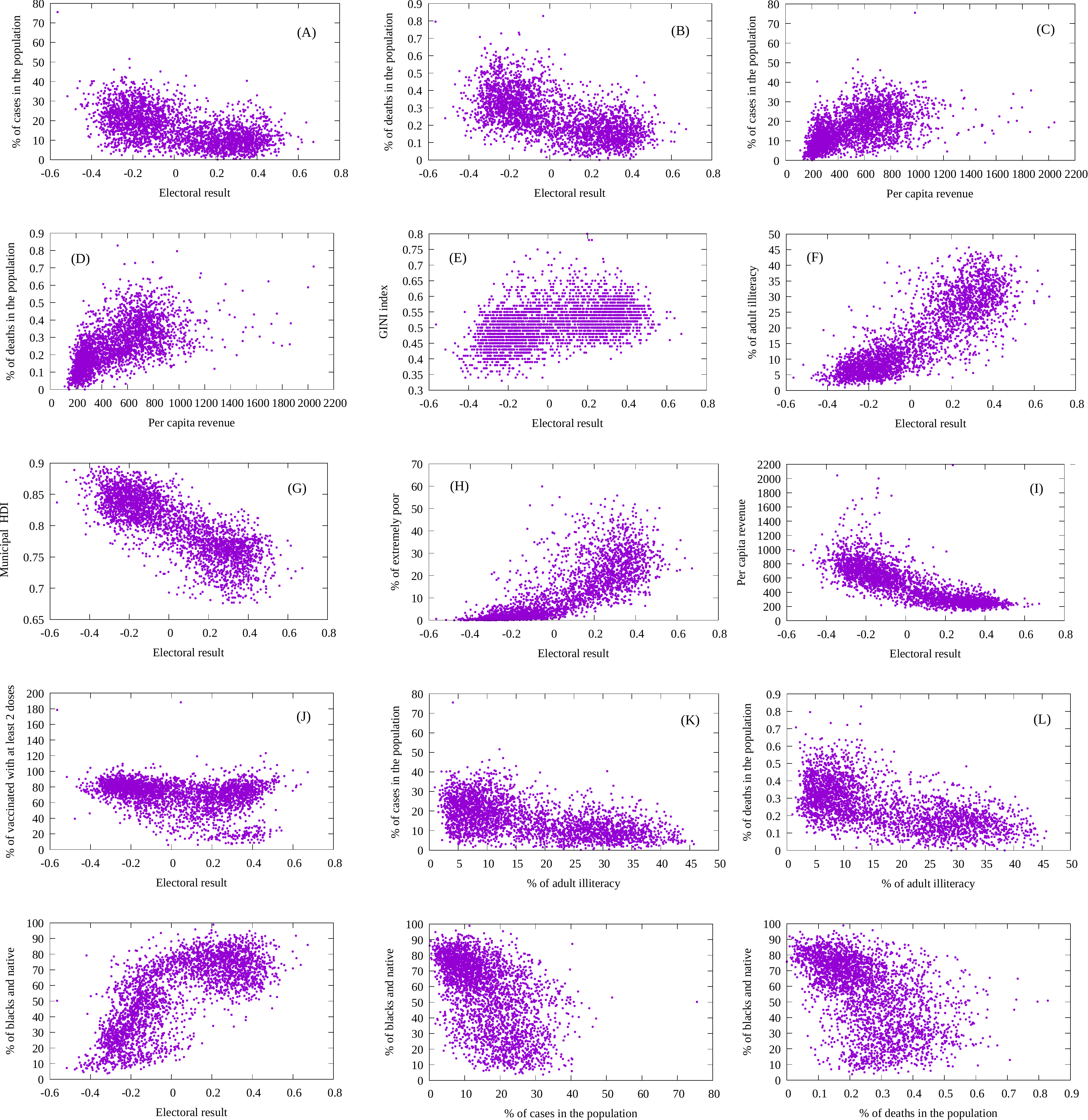}}
\end{center}
        \caption{Scatter plots displaying for a few pairs of variables in Table~\ref{tab1}.\label{scater}}
\end{figure}

The scatter plots in Fig.~\ref{scater} provide visual representations for some pairs of variables in Table~\ref{tab1},
where relevant correlations are visible. The Spearman correlation coefficient $r_s$ values for all pairs of data values
are shown in Fig.~\ref{heatspear} (the statistical significance analysis is provided in the supplemental material),
demonstrating a clear tendency for municipalities with a majority voting for Bolsonaro to have a worse pandemic outcome, i.e.,
the higher the proportion of votes in Bolsonaro, the higher the attack rate and the proportion of deaths in the population
(see also Figs.~\ref{scater}A, \ref{scater}B, and~\ref{scater}C). In the same way. A higher
human development index and a higher average income also correlate with a vote for Bolsonaro.
On the other hand, the left-wing candidate Haddad received a higher proportion of votes in municipalities
with higher adult illiteracy, higher inequality as measured by the GINI index,
a higher proportion of extreme poverty, and a higher proportion of black and indigenous populations. Those are signs
of a highly segregated social structure that translates not only in electoral preferences but also in
the expression of pro-social behaviors relevant to the mitigation of a crisis, such as the current pandemic.

\begin{figure}[ht]
\begin{center}
        \scalebox{0.9}{\includegraphics{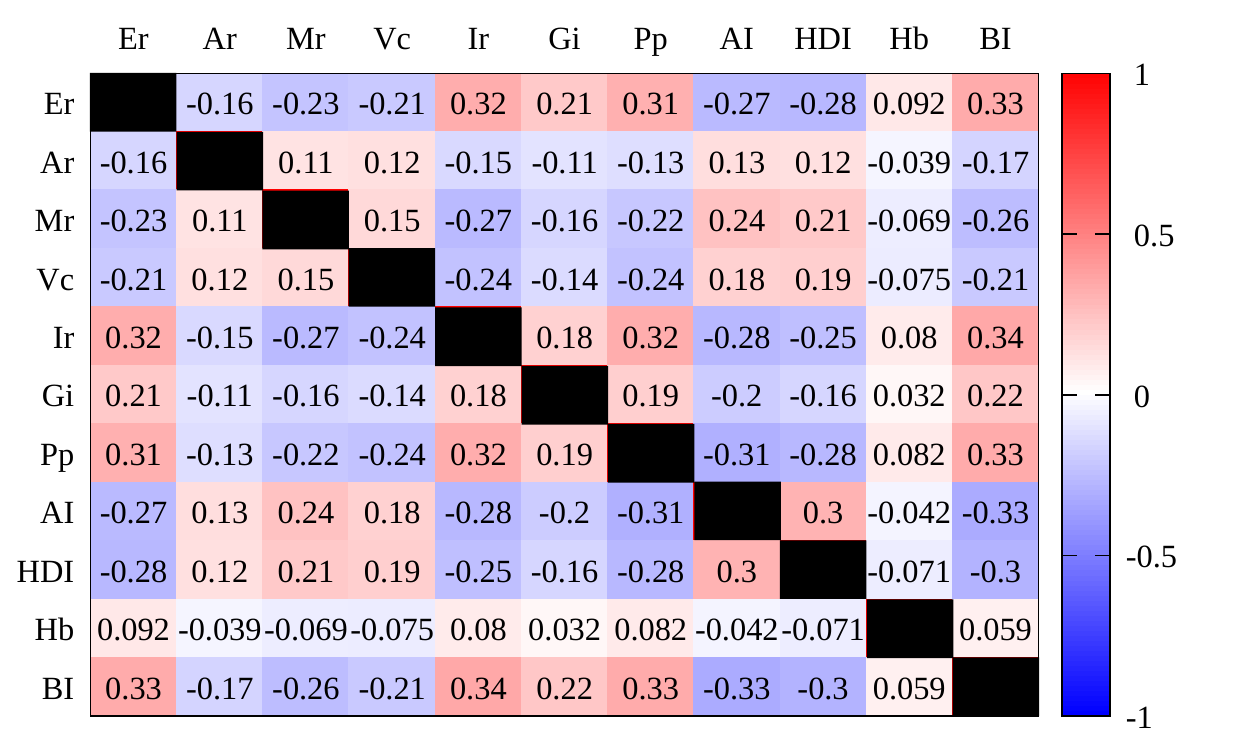}}
\end{center}
        \caption{Heatmap and values of Spearman correlations $r_s$ between the variables in Table~\ref{tab1}.\label{heatspear}}
\end{figure}

The correlation between electoral results and vaccination needs a closer look.
In the scatter plot in Fig.~\ref{scater}J, no tendency for higher or lower vaccination rates can
be detected except for a number of municipalities in
the lower right portion of the plot (a lower vaccination rate and higher voting in Haddad).
By inspecting the vaccination data, we observe that
there are 128 municipalities with less than 30 percent of fully vaccinated individuals,
with 140 of them in the state of Ceará. This is either due
to a local problem in data gathering or some other issue specifically present in this state.
With this observation, we conclude that there is no clear
difference in vaccination rates between municipalities that voted for either candidate.

The current analysis shows a positive correlation between the
proportion of black and indigenous populations and data related to economic inequality.
Nevertheless, both the attack rate (the proportion of cases in the population) and the mortality rate from COVID-19
have a significant negative Spearman correlation with the proportion of black and indigenous populations,
which also tended to vote for the left-wing candidate.
It is noteworthy that localities where Haddad obtained the majority of votes are also those that
have a higher budget in the public health sector per inhabitant (by 2021), despite having a lower average income,
which may be related to local governments (mayors and and state governors) being more prone to finance public
structures and not having to rely on a very present private sector.

\subsection{Excess deaths caused by denialism}

\begin{figure}
\begin{center}
        \scalebox{0.9}{\includegraphics{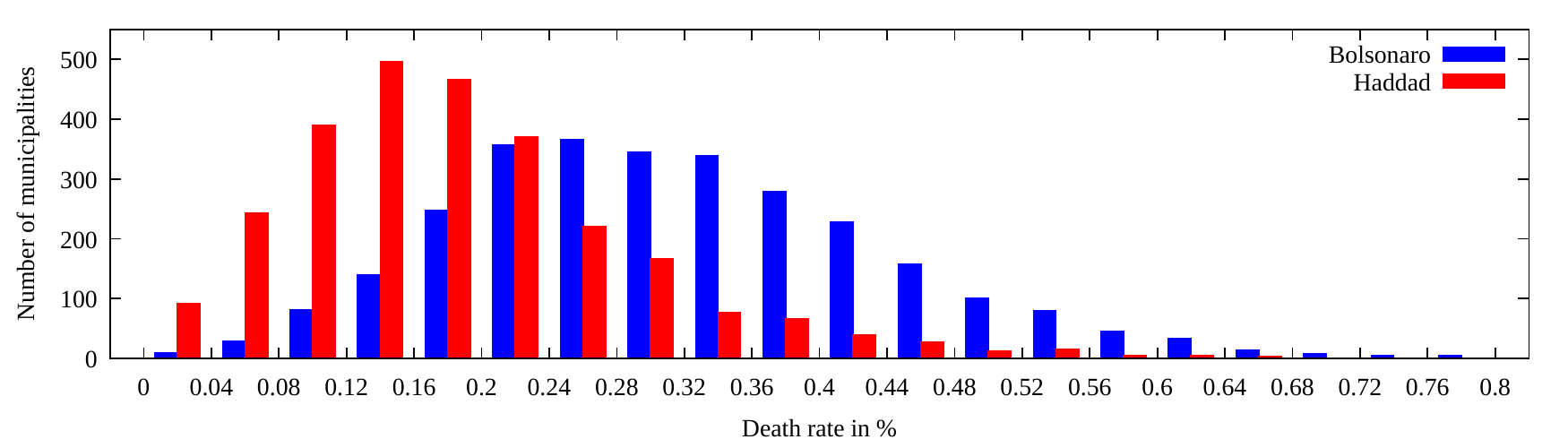}}
\end{center}
        \caption{Number of Brazilian municipalities according to total mortality by COVID-19 at September,12 2021
	and according to the winner candidate at the second round of the Brazilian presidential election.\label{fig3}}
\end{figure}

In order to show that COVID-19 denialism is related to polytical preferences, we show in Fig.~\ref{fig3}
the histograms of the number of Brazilian municipalities according to the official death rate and
the winning candidate. We observe that the death rate for
COVID-19 is significantly higher in those places where Bolsonaro won than in those places where Haddad won
that cannot be dismissed as a mere coincidence.
As a first estimate of the excess deaths caused by far-right nationalism in Brazil, we suppose that municipalities with
similar populations could have achieved the same death rate if mitigation measures
were properly implemented. We thus divide all municipalities into 20 groups according to increasing population,
each with 277 municipalities. The death rate obtained for each group determined from a weighted average with respect to
population is shown in Table~\ref{mortpop}. The expected number of deaths in each population group is then obtained by
supposing that the death rate in Bolsonaro-prone municipalities were the same as in those where Haddad won.
This results in a number of $524\,501$ total deaths, which is to be compared to the official number of $683\,407$ deaths,
pointing to the fact that $158\,906$ deaths occurred due to either believing in the far-right speech of Bolsonaro and his staff or
from a lack of proper measures from local authorities following the directives of the central government.
The estimate is based on a few assumptions, but we stress that it would be almos impolssible to  explain
the significant difference in the oberved death rates than resorting to the consequences of denialism.

\begin{table}[ht]
        \begin{center}
\begin{tabular}{ccc}
        \hline
	Population range & Avg. death rate \% & Avg. death rate \% \\
                         & (Bolsonaro) & (Haddad) \\
        \hline\hline
	 807 – 2470 & 0.309 & 0.218\\
	 2470 - 3162 & 0.314 & 0.217\\
	 3163 - 3853 & 0.298 & 0.205\\
	 3865 - 4596 & 0.303 & 0.197\\
	 4596 - 5366 & 0.301 & 0.204\\
	 5371 - 6260 & 0.304 & 0.188\\
	 6266 - 7291 & 0.290 & 0.182\\
	 7294 - 8494 & 0.309 & 0.191\\
	 8501 - 10076 & 0.317 & 0.179\\
	 10077 - 11573 & 0.315 & 0.179\\
	 11595 - 13443 & 0.294 & 0.171\\
	 13478 - 15611 & 0.294 & 0.170\\
	 15614 - 18210 & 0.326 & 0.174\\
	 18218 - 21334 & 0.328 & 0.170\\
	 21356 - 25596 & 0.307 & 0.166\\
	 25600 - 31815 & 0.324 & 0.172\\
	 31824 - 41557 & 0.336 & 0.169\\
	 41595 - 61190 & 0.337 & 0.192\\
	 61255 - 117481 & 0.340 & 0.201\\
	 118583 - 12228009 & 0.383 & 0.292\\
        \hline
\hline
\end{tabular}
        \end{center}
        \caption{Average death rate according to population range. Average values were obtained as a weighted average according
	to population in each municipality.\label{mortpop}}
\end{table}

Different estimates can be obtained by grouping the municipalities according to different indices and rates given in Table~\ref{tab1}.
We also separate the municipalities according to population size but now only in three groups:
up to 10000 inhabitants (2486), from 10001 to 50000 (2396), and more than 50000 (682), with
the number between parenthesis the respective number of municipalities.
We then dividei each such groups based on a range of values for each index, each one again divided according to the
presidential electoral results. We then comipute in each one the population-weighted average
death rate by COVID-19.
The excess death toll is then determined, as in the previous case, by applying the death toll from those municipalities
that voted for Haddad to those that voted for Bolsonaro. We first consider the per-capita public health budget ($H_b$).
The results are shown in Table.~\ref{tab3}, with an estimated total of $494\,955$ deaths, plus $11$ deaths in municipalities
with no data on $H_b$, and therefore
an estimated number of deaths due to denialism of $188\,444$.
Other estimates obtained by a smiliar procedure using other indiced are given in
Tables~S2--S6 of the Supplemental Material, and summarized in Table~\ref{tabexcess}.
Although the results vary significantly, they all point to an expressive number of lives lost to
disinformation and a lack of proper care by authorities following a denialist stance.
Other common factors such as lack of proper health structure
and of public campaigns to inform the population also result in an unecessary loss of lives,
and the estimates obtained here are thus a low bound.

\begin{table}[ht]
        \begin{center}
\begin{tabular}{ccccc}
        \hline\hline
	Population range & Range of values & Mortality \% & Mortality \% & Expected deaths \\
	&  & (Bolsonaro) & (Haddad) & \\
        \hline\hline
	 $\leq10000$  & $H_b\leq 100$ (0) & -- & -- & --\\
	 &      $100<H_b\leq200$ (0) & -- & -- & --\\
	 &      $200<H_b\leq500$ (1) & -- & 0.226 & 17\\
	 &      $H_b>500$ (2482) & 0.165 & 0.160 & 20671\\
	 \hline
	 $10001$--$50000$  & $H_b\leq 100$ (4) & 0.399 & 0.404 & 413\\
	 &      $100<H_b\leq200$ (9) & 0.336 & --  & 859\\
	 &      $200<H_b\leq500$ (289) & 0.307 & 0.166 & 17013 \\
	 &      $H_b>500$ (2094) & 0.329 & 0.134 & 55919\\
        \hline
	$>50000$  & $H_b\leq 100$ (99) & 0.392 & 0.312 & 246892 \\
	 &      $100<H_b\leq200$ (158) & 0.363 & 0.217 & 72387\\
	 &      $200<H_b\leq500$ (344) & 0.358 & 0.211 & 66709\\
	 &      $H_b>500$ (82) & 0.366 & 0.261 & 14071\\
\hline
 Total & & & & 494955\\
	\hline\hline
\end{tabular}
        \end{center}
	\caption{Estimated deaths by COVID-19 by considering the mortality in municipalities that voted in majority
	for Bolsonaro (far right) would be the same as in the municipalities that voted in Haddad (left-wing)
	for each sub-division in population and per-capita public Health budget $H_b$ for the intervals $<100$, $101$ to $200$,
	$201$ to $500$ and $>500$ in Brazilian Reais. The numbers of municipalities in each class are given
	among parenthesis.\label{tab3}}
\end{table}

\begin{table}[ht]
        \begin{center}
\begin{tabular}{lcc}
        \hline\hline
	 Index & Total estimated & Total excess\\
	       & deaths & deaths\\
        \hline\hline
	Population &  524501 & 158910\\
	Health budget ($H_b$) & 494966 & 188444\\
	GINI coefficient ($G_i$) & 493329 & 190082\\
	Extreme Poverty rate ($P_p$) & 558125 & 125286\\
        Human Development Index ($HDI$) & 577316 & 106095\\
	proportion of black and indigenous population ($BI$) & 541096 & 142315\\
        \hline\hline
\end{tabular}
        \end{center}
        \caption{Estimated deaths by COVID-19 caused by denialism in Brazil by grouping municipalities according
	to different social and economic indices. The official total toll by September 23 is of 683411 deaths.\label{tabexcess}}
\end{table}

\section{Discussion and conclusions}

Behaviors for preventing the dissemination of the SARS-CoV-2 virus,
based on scientifically sound and effective measures~\cite{Rocha2021,Commission2022},
such as personal hygiene measures, mask wearing, social distancing, avoiding crowds, and quarantining when ill,
are linked to trust in government~\cite{Han2020}.
President Bolsonaro made numerous speeches in which he questioned the scientific evidence
on the efficacy of such non-pharmaceutical interventions (seei~\cite{BContr}).
At the height of the pandemic, with more than four thousand deaths in a single day, no awareness campaigns
were deployed by the central federal government (and barely timid ones by state governments)
to properly guide the general population. In fact, considerable effort was expended in calling into
question the efficacy of all valid mitigation measures, which were implemented
in the majority of countries where public policies are dictated democratically. Nevertheless, his speeches
didn't have a homogeneous reach across the country, as shown by the data presented here.
Municipalities with lower economic indices and a higher proportion of marginalized people
tended to vote left, while those with higher indices tended to vote right.
Counterintuitively, prossocial behavior at the municipality level is a proxy
for worse economic and social indicators, as richer localities fared worse against the pandemic,
despite having more acces to private health facilities dur to the average hogher income.
The outcomes of the COVID-19 pandemic demonstrate the intensity of political,
social, and economic relations in Brazilian society,
either in a democratic or an authoritarian framework~\cite{Avritzer2021}.

The authorities in the Brazilian central government failed to recognize how serious the
situation was and even propagated false beliefs and irresponsible attitudes, such as denying the existence of efficient therapies
in favor of scientifically proven inefficient medications and even delivering public speeches against mass vaccination. Therefore,
it comes as no surprise that the burden of the pandemic was higher in those municipalities where authorities sympathized with and supported
such policies, as a sad consequence of a divided society. The mandatory use of masks, which
was implemented by the state governments despite the opposition of the federal government, and other non-pharmaceutical measures
were dependent on the willingness and conformity of local authorities, as well as individual behavior, which was
closely influenced by the dominant beliefs.

The consequences of denying scientifically established facts can already be seen in public data audited
by the Brazilian external control authority, the Federal Court of Auditors ({\it Tribunal de Contas da Unãio})~\cite{tcu}.
Only on October 19, 2020, the Center of Government,
composed of the Presidency of the Republic, the General Secretariat of the Presidency, and the Chief of Staff,
the highest decision-making level in the country, created a Crisis Committee for the supervision and monitoring of the impacts of
the health crisis caused by the COVID-19 ándemic. The committee was formed more than a year after the first case was detected in the country.
However, as pointed out by the auditors, there were no general diagnoses that report on the sanitary and health
situation in each Brazilian state, nor have guidelines been outlined so that managers across the country could rely
to guide their decisions. It also emphasizes that no scenarios were developed, nor were the risks that loom over the country.
The court of accounts pointed to the central government hesitation to assume its constitutional
role, namely to carry out actions to fight and mitigate the pandemic.
It also reported omissions from both by the Center of Government and the Committee established
by it to deal with the serious health crisis. This hesitancy influenced the behavior of many local
authorities aligned with the far-right policies of the central government, as well as followers
in the general population who believed and acted on the Brazilian president's denialism speech,
resulting in far more deaths than would have occurred if scientific-based policies had been implemented
at the start of the pandemic, as demonstrated in the current work. Indeed, Brazil
is the second country with the highest number of deaths in the world, just after the United States,
and is in the $20^{\rm th}$ place in deaths by millions of inhabitants among 230 countries with
available data~\cite{Worldometer2022}.

We conclude by stressing that the election of President Bolsonaro in 2018,
which impacted the failure to properly fight the dissemination of the SARS-CoV-2 virus,
is not the cause of many mishaps in Brazil but a symptom of an underlying contradictory social and economic structure that,
among other consequences, hinders the development and evolution of a stable, modern, democratic, and inclusive society in Brazil.

\bibliography{refs.bib}

\end{document}